\documentclass[prl,twocolumn,showpacs,preprintnumbers,amsmath,amssymb]{revtex4}

\usepackage{graphicx}
\usepackage{dcolumn}
\usepackage{bm}
\usepackage{hyperref}
\usepackage{breakurl}

\begin{document}


\title{High accuracy correction of blackbody radiation shift in an optical lattice clock}

\author{Thomas Middelmann}
\author{Stephan Falke}
\email{stephan.falke@ptb.de}
\author{Christian Lisdat}
\author{Uwe Sterr}
\affiliation{Physikalisch-Technische Bundesanstalt (PTB), Bundesallee 100, 38116 Braunschweig, Germany}

\date{\today}

\begin{abstract}

We have determined the frequency shift that blackbody radiation is inducing on the $5s^2$ $^1$S$_0$ -- $5s5p$ $^3$P$_0$ clock transition in strontium. 
Previously its uncertainty limited the uncertainty of strontium lattice clocks to $1\times10^{-16}$.
Now the uncertainty associated to the blackbody radiation shift correction translates to $5\times 10^{-18}$ relative frequency uncertainty at room temperature.
Our evaluation is based on a measurement of the differential dc-polarizability of the two clock states and on a modeling of the dynamic contribution using this value and experimental data for other atomic properties. 
\end{abstract}


\maketitle
The performance of optical clocks promises a large variety of benefits. Ultimately the definition of the SI unit for time and frequency could be reworded to exchange the microwave transition as reference by an optical one. At present the realization of the second would be improved in accuracy by more than an order of magnitude \cite{ros08}; its stability could be increased even by several orders of magnitude \cite{jia11, tak11, hag12}. Already without a changed definition, optical clocks serve due to their accuracy to probe the universe in real time for temporal variations of the fundamental constants \cite{ros08,pei04,dzu09} like the fine structure constant $\alpha$ or coupling of fundamental constants to forces like e.g. gravity \cite{bla08}. In combination with their outstanding stability they are also considered to be the key instruments that could enable new measurement opportunities like relativistic geodesy \cite{kat11,bje85} with high temporal resolution as supplement to established gravimetric techniques by comparing gravitational potentials directly.

In the quest for best clocks, lattice clocks offer highest stability \cite{jia11, tak11, hag12} but suffer from blackbody radiation shifts \cite{cam08b,fal11,she12}.
Among these, strontium clocks are pursued in several experiments worldwide \cite{cam08b, bai08, fal11, yam11}. Their systematic uncertainty now reaches the $1\times10^{-16}$ level  and is thus significantly below that of primary frequency standards used for frequency measurements. 
The correction for the shift by the blackbody radiation (BBR) typically dominates the uncertainty budget of the strontium clocks. 
Besides uncertainty due to the temperature measurement, at present the uncertainty of the atomic response to BBR at room temperature contributes with $7 \times 10^{-17}$ to the uncertainty.  
The obvious strategy to improve clock accuracy by reducing the environmental temperature requires involved apparatuses. Instead in most clocks accurate corrections of the BBR shift are applied.
This correction is so far only known from atomic structure calculations  \cite{por06,por08} and lacks experimental verification.  

As the frequency of room temperature BBR is very low, measurements of the  difference of the dc polarizabilities of the two clock states greatly improves the knowledge about the BBR shift. In case of an optical frequency standard with neutral ytterbium, recent experiments demonstrated a significant reduction of uncertainty of the BBR shift correction by measuring the frequency sensitivity of the clock transition to a dc electric field \cite{she12}. Similar approaches have been pursued for Cs clocks \cite{sim98, rob10}.

\begin{figure}[t]
\includegraphics[width=\columnwidth]{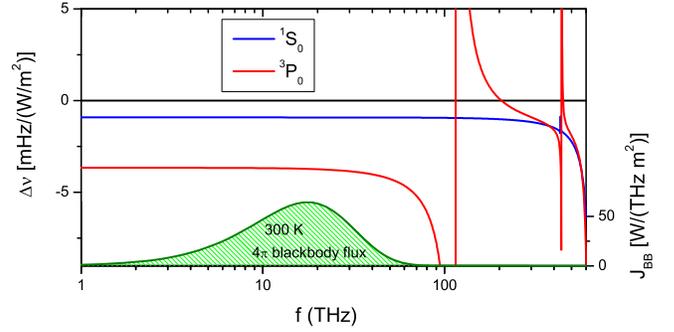}	
\caption{ac-polarizabilty of the two clock states $^1$S$_0$ and $^3$P$_0$ as a function of the frequency and room temperature blackbody spectrum $J_{\rm BBR}$. 
The resonance structure in the $^3$P$_0$ polarizability is due to 115~THz $5$s$4$d$~^3$D$_1$--$^3$P$_0$ transition. The magic wavelength lattice is at the crossing of both ac-polarizability curves at 369~THz.}
\label{fig:level}
\end{figure}

However, in both strontium and ytterbium, more than the differential static polarizability $\Delta \alpha$ needs to be considered at the level of $10^{-16}$ relative frequency uncertainty due to low-frequency lines (Fig.~\ref{fig:level}). 
In the BBR shift a dynamical correction $\Delta \nu_{\rm dyn}(T)$ has to be applied \cite{por06}:
\begin{equation}
	\Delta\nu_{\rm BBR} = -\frac{1}{2h}\Delta \alpha\langle E^2 \rangle_T + \Delta \nu_{\rm dyn}(T) ,
	\label{eq:nuBBR}
\end{equation}
where $\langle E^2 \rangle_T \approx \left(8.319~{\rm V}/{\rm cm}\right)^2\times\left(T/300~{\rm K}\right)^4$ is the mean squared electric field of the BBR radiation of temperature~$T$. To first order, $\Delta \nu_{\rm dyn}(T)$	scales with $T^6$.
In case of strontium it contributes to the relative frequency shift due to BBR at a level of $3\times 10^{-16}$ \cite{por06,por08}.

In this work we reduce the uncertainty of the BBR frequency shift coefficient of the 429~THz $(5{\rm s}5{\rm p}) ^3{\rm P}_0$ -- $(5{\rm s}^2) ^1{\rm S}_0$ clock transition  of Sr in two steps: First, through a measurement of the differential dc-polarizability $\Delta\alpha$ to provide a high precision experimental value for the largest fraction of the shift. 
Second, we use this observable along with several others to model the dynamic BBR shift, i.e. $\Delta \nu_{\rm dyn}$. From the improved model of the response to BBR, the so far leading contribution to the uncertainty of Sr clocks can now be reduced by an order of magnitude to $5\times10^{-18}$ at room temperature.

To measure $\Delta\alpha$ we equipped our strontium lattice clock apparatus with a precision capacitor and a moving optical lattice setup \cite{mid12} to transfer atoms from the loading to the interrogation region. 
Samples of few $10^4$ $^{88}$Sr atoms at around 2~$\mu$K are created by Zeeman slowing and laser cooling in a two-stage magneto-optical trap (MOT) \cite{leg09} and trapped in the horizontal one dimensional lattice with a waist radius of 65~$\mu$m and trap frequencies of 70~kHz axial and 200~Hz radial at a trap depth of 9~$\mu$K. 

\begin{figure}[t]
\includegraphics[width=\columnwidth]{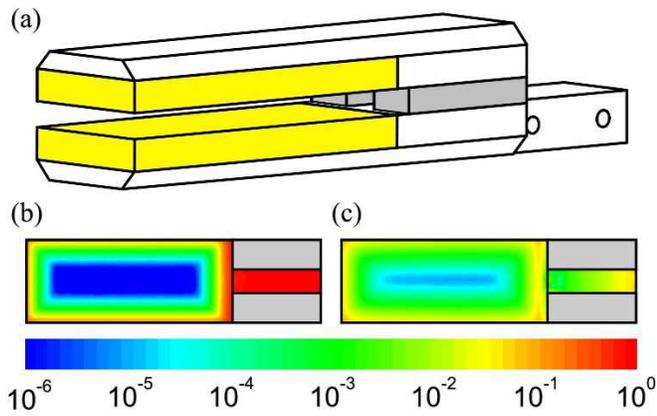}	
\caption{Precision capacitor. Part (a) shows a sketch to scale. The yellow areas depict the gold plated electrodes. Part (b) shows the deviation of the electric field from an infinite capacitor with $U/d$, when voltages $\pm U/2$ are applied. Part (c) shows the $z$ component of the stray field $\vec{E}_0$ leaking into the capacitor on a plane 1~mm above the center when both electrodes are on the potential $U/2$. Both fields from finite element calculations are normalized to $U/d$.}
\label{fig:cap}
\end{figure}

To maintain the trap depth throughout the transport of atoms over several Rayleigh ranges of the focused lattice beams, the waist of the lattice beams is moved together with the interference pattern. 
This is achieved by moving the beam shaping optics and retro-reflection unit with translation stages. 
This setup allows to move the atoms from the MOT region into the capacitor and, after interrogation by the clock laser, back out for detection by laser induced fluorescence on the $(5{\rm s}5{\rm p}) ^1{\rm P}_1$ -- $(5{\rm s}^2) ^1{\rm S}_0$ transition. We detect also the atom number in the $^3{\rm P}_0$ state by repumping them to the $^1{\rm S}_0$ state to reduce atom number fluctuation related noise. To avoid a first order Doppler shift during the interrogation of the clock transition due to residual motion of the stages we stabilize the path length between clock laser and atom position represented by the retro-reflection mirror of the lattice \cite{fal12}.

The capacitor consists of two Zerodur plates, separated by two optically contacted Zerodur gauge blocks of 5~mm height (see Fig.~\ref{fig:cap}) that was determined interferometrically using three wavelengths \cite{fra10}. The plates are partially coated with semitransparent gold layers (20~nm) and underlying contact layers of aluminum (5 nm) to allow for an interferometric determination of the electrode separation before installation. The measured surface flatness of the plates leads to a correction of $-153(28)$~nm (Tab.~\ref{tab:uncert1}) at the position of the atoms. Here, we observe locally an angle between the electrodes of 5~$\mu$rad, which corresponds to the measured parallelism of the gauge blocks' surfaces and the flatness of the electrode plates. After assembling the capacitor the field plate separation was confirmed by an independent direct interferometric measurement. As the plate separation was measured before the capacitor was brought into the vacuum chamber, we also include contributions as relaxation under vacuum, bending of the plates under the influence of gravity or thermal expansion. The conductive surface of $28~{\rm mm} \times 67{\rm~mm}$ ensures field homogeneity of better than $10^{-6}$ over a large area (Fig.~\ref{fig:cap}).

\begin{table}[t]
\begin{tabular}{lr@{}lr@{}l}
\hline
Source			& \multicolumn{2}{c}{correction}		& \multicolumn{2}{c}{uncertainty}\\
						& \multicolumn{2}{c}{in nm}					& \multicolumn{2}{c}{in nm}\\
\hline
\hline
Gauge block height  & --&												& $11$ & $.3$\\
Contacting 					& 0&												& 11&.3\\
Coating 						& $-50$&										& 15&\\
Parallelism 				& $-153.$&1									& 27&.7\\
Position						& 0&												& 10&\\
Bending of plates  	& 0&											  & 13&.1\\
Electrostatic forces	& 0&										& 0&.04\\
Compression by air pressure	& 2&.8 								& 0&.1\\
Temperature gauge blocks 	& 0&									& 2&.5\\
Temperature field plates	& 0&									& 6&.0\\
\hline 
\textbf{Separation}	& $5\,001\,644$ &$.2$
& \hspace{5ex}$39$ &$.5$\\
\hline 
\end{tabular}
\caption{Uncertainty budget for the capacitor plate separation.}
\label{tab:uncert1}
\end{table}

Voltage is supplied to the electrodes by two electric wires each, connected to the electrodes at points with maximum separation to also allow for measurements of residual resistance of the connections. 
Independent voltages of up to 700~V can be applied to the electrodes from two precision voltage supplies (Fluke 415B). 
The voltage difference is monitored by a precision voltmeter (HP 3458A calibrated via a Fluke 5720A calibrator to a Josephson voltage standard), with an averaging time of 1000 power line cycles (50~Hz). The electrodes are connected to the voltage supplies with reed relays, which allow for inverting or discharging the capacitor without interrupting the voltage measurement.

We measure the field induced shift of the atomic resonance in $^{88}$Sr in three field configurations (applied field, inverted field and no field) by interleaving three stabilizations \cite{deg05a, fal11} of the clock laser to the atoms.  
Information on the shift induced by the capacitor $\vec{E}_{\rm cap}$ and small bias fields $\vec{E}_0$ is contained in the three offset frequencies $\nu_i$ ($i= \uparrow, \downarrow, 0$) to the reference cavity of the clock laser. 

\begin{figure}[t]
\includegraphics[width=\columnwidth]{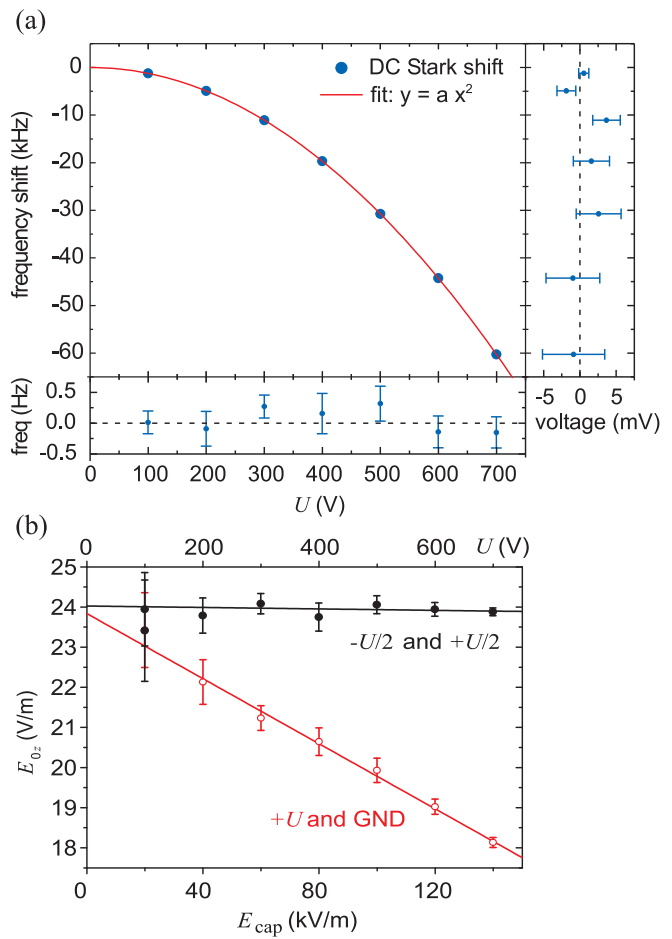}	
\caption{Measured differential dc Stark shift. (a) shows the measured data and a fitted parabola along with the residuals. (b) shows the measured bias field component in $z$-direction.}
\label{fig:para}
\end{figure}

As the observed shift is proportional to $\left( \vec{E}_{\rm cap}+\vec{E}_0 \right)^2$, the dc-Stark shift due to $\vec{E}_{\rm cap}$ alone is given by
\begin{equation}
\nu_{\rm dc}=(\nu_\uparrow + \nu_\downarrow)/2 - \nu_0,
\label{eq:shift}
\end{equation}
provided changes in $\vec{E}_0$ are slow compared to the cycle times.
The data of a typical experimental run are shown in Fig.~\ref{fig:para} with residuals from a parabolic fit.

The component of a small bias field $E_{0z}$ along the direction of the applied field $\vec{E}_{\rm cap}$ is given by 
\begin{equation}
\nu_\uparrow - \nu_\downarrow \propto \vec{E}_{\rm cap} \cdot \vec{E}_0 = E_{\rm cap} E_{0z}.
\end{equation}
The bias field $E_{0z}$ can be explained as the sum of patch fields independent of the applied field and stray fields. 
Patch fields can occur e.g. due to a difference in the work function of the gold coatings, which depends on the crystalline structure and can have a magnitude of up to 500~meV \cite{rob06,mic77}. They explain the part of $E_{0z}$ that is independent of the applied voltage (Fig.~\ref{fig:para}).
However, differences in the work function do not explain the voltage dependence of $E_{0z}$. As this behavior may indicate residual differences between the voltage of the capacitor and the measured voltage, it was crucial to understand it:
The system can be described by three electrodes (two capacitor electrodes at potentials $U_1$ and $U_2$ and the surrounding vacuum system on ground potential). 
Any field in this model can be written as the sum of the antisymmetric configuration with electrodes at $\pm(U_1-U_2)/2$ and a stray field with both electrodes at the average potential $(U_1+U_2)/2$. 
This stray field $\propto U_1+U_2$ is contributing to $E_{0z}$ if the atoms are not probed exactly on the symmetry axis of the capacitor (Fig.~\ref{fig:cap} part~(c)).  
Simulations confirm that an offset of 1~mm above the capacitor's symmetry plane with a shift of 1~mm to one side explains the observed dependence on the applied voltage, which is compatible to our effort on positioning.

This stray field explains the voltage dependence of $E_{0z}$ and thus Eq.~\ref{eq:shift} can be applied to determine $\nu_{\rm dc}$ (neglecting an insignificant quadratic contribution from $\vec{E}_0$). No significant time dependence of $E_{0z}$ is observed. Along with measured discharging curves of the electrodes and measurements of the resistance of the capacitor plates against ground potential and between feed lines to one plate we conclude that the measured voltage $U$ is determining $E_{\rm cap} = U/d$.  

Several parabolas as in Fig.~\ref{fig:para}~(a) were measured and $\Delta\alpha$ was derived for each voltage. 
We confirmed that neither alternating the order of the three interrogations nor adding a second zero field cycle did cause any systematic effect. 
No significant day-to-day variation of $\Delta\alpha$ was observed. 
To properly account for different uncertainty contributions of the voltmeter we first averaged the $\Delta\alpha$ values for each applied voltage.
In a second step all these values were averaged with weights according to the uncertainty. We obtain the differential dc-polarizability $\Delta\alpha = 4.07873(11)\times10^{-39}~{\rm Cm}^2/{\rm V}$. An influence of the hyperpolarizability scaling with $E^4$ was investigated by fitting a parabola of type $a+bE^2$ to the $\Delta \alpha$ obtained in the first averaging step. The difference between $a$ and the weighted average of $\Delta \alpha$ was taken as uncertainty contribution for the hyperpolarizability.
A complete list of uncertainty contributions to $\Delta\alpha$ is given in Tab.~\ref{tab:uncert2}.

\begin{table}[t]
\begin{tabular}{lr@{}l}
\hline
Source & fractional & ~contribution 
\\
& \multicolumn{2}{c}{ $(10^{-6})$}\\
\hline
\hline
Separation														&	$15$ 	&	$.8$\\
Voltage $U$\\
\hspace{3ex} Realization							&	18		&	.5\\
\hspace{3ex} Charging of capacitor 		&	5			&	.0\\
\hspace{3ex} Inverting, drift 				&	2			&	.0\\
\hspace{3ex} Residual voltage divider	&	0			&	.3\\
Fringe and stray fields 							&	2			&	\\
Time varying bias fields							& 0			&	.3\\
Hyperpolarizability										& 12		&	\\
Statistical uncertainty $\nu$, $U$		&	4			&	.0\\
\hline 
Total uncertainty 						& 28	&	\\
\hline 
\end{tabular}
\caption{Uncertainty budget of measurement of the differential dc polarizability $\Delta\alpha$.}
\label{tab:uncert2}
\end{table}

In comparison the atomic structure calculations of Porsev and coauthors \cite{por06} give $\Delta\alpha = 4.305(59)\times10^{-39}~{\rm Cm}^2/{\rm V}$; a difference of more than $3\sigma$ to our value. 
This is still within the typical range of deviation between calculated and experimental values, e.g. lifetimes in \cite{por08}.

As discussed in the introduction, for improving strontium lattice clocks to the $10^{-18}$ uncertainty range, the dynamic shift contribution $\Delta \nu_{\rm dyn}$ (Eq.~\ref{eq:nuBBR}) needs to be determined. 
We calculated its value by first determining the ac-polarizabilities $\alpha_i(\omega)$ of the clock states ($i=\rm{e,g}$) (Fig.~\ref{fig:level}) using Einstein coefficients $A_{ki}$: 
\begin{equation}
\alpha_i(\omega) = 2\pi \epsilon_0 c^3 \sum_{k}
\frac{2 J_k+1}{2 J_i+1} \frac{A_{ki}}{\omega_{ik}^2(\omega_{ik}^2-\omega^2)},
\label{eq:alpha}
\end{equation}
and then integrating the differential ac-Stark shift over the Planck distribution \cite{mid11}. Here $\omega_{ik}$ denotes the angular frequency of the transition $k\rightarrow i$ and $J_i$ and $J_k$ the angular momenta.

We used our measured static polarizability in combination with other observables as magic wavelength, atomic lifetimes, Thomas-Reiche-Kuhn sums \cite{con53}, and ac Stark shifts of the clock laser field \cite{bai07} to adjust the $A_{ki}$ with the largest contribution in a least square fit (see Supplement). 
In this way we include the best experimental knowledge to improve $\Delta \nu_{\rm dyn}$. 

In order to obtain an uncertainty estimation for $\Delta \nu_{\rm dyn}$, we performed 2000 fit runs to synthetic datasets, i.e., sets of observables varied at random within their respective uncertainties \cite{deg04}.
From the Monte-Carlo simulation (see Supplement) we find that the dynamic BBR shifts of the two clock levels at $T_0 =300$~K are $-150.4$~mHz and $-2.8$~mHz for exited and ground state.

In summary, the clock transition resonance frequency is shifted by BBR of temperature $T$ by 
\begin{equation}
\Delta\nu(T) = \Delta\nu_{\rm stat} \left( \frac{T}{T_0} \right)^4 + \Delta \nu_{\rm dyn} \left[ \left( \frac{T}{T_0} \right)^6 + \mathcal{O}\left( \frac{T}{T_0}\right)^8 \right].
\label{eq:corr}
\end{equation}
with 
$\Delta \nu_{\rm stat} = -2.13023(6)$~Hz 
and 
$\Delta \nu_{\rm dyn} =  -147.6(23)$~mHz.
Compared to the corrections used so far \cite{por08} of $2.354(32)$~Hz at $300$~K, applying our correction shifts the clock frequency by a fraction of $-1.8\times10^{-16}$. 
While this is on the order of the uncertainties of the best Sr lattice clocks, it is below the uncertainties of the measured frequencies. 
In the course of this work a numerical error in the calculation of the dynamic corrections in reference \cite{por06} was identified \cite{der12}. 
This error amounts to $-53$~mHz, while this new study changes the correction due to the static part by $118$~mHz and the dynamic part by $11$~mHz.

These corrections immediately apply to $^{88}$Sr. To obtain the corrections for $^{87}$Sr we have repeated the calculation according to Eq. \ref{eq:alpha} using published data on isotopic shift and hyperfine structure \cite{asp91, ans86, let07a, lor83, buc85, eli83} (see Supplement) and using a mass scaling of the transition dipole matrix elements \cite{wie09a}. 
All published transition isotope shifts between $^{88}$Sr and $^{87}$Sr lie in a range of up to 150 MHz. Thus for the calculation for lines with unknown isotope shift a conservative estimate of 150 MHz was assumed.
With these data we obtained a fractional change of $\Delta \alpha$ of $-1.2\times10^{-6}$ and of $\Delta \nu_{\rm dyn}$ of $-3.4\times 10^{-6}$.
Thus the correction (Eq.~\ref{eq:corr}) also apply to $^{87}$Sr well within their uncertainties.  

In conclusion, a measurement of the differential dc-polarizability of the strontium clock transition in combination with a model for the frequency response of the atoms to BBR enabled us to derive an improved correction of the BBR shift and estimate of its uncertainty. 
With this study, corrected frequencies of a number of high precision measurements of the clock transition may now be calculated \cite{cam08b, bai08, fal11, yam11}. The corrected frequencies still agree well as the applied correction is smaller than typical uncertainties of realizations of the SI second by Cs clocks.
Now the shift correction coefficient leads to an uncertainty of $5\times10^{-18}$ at room temperature, where a probably achievable temperature uncertainty of 80~mK would add the same uncertainty.

To further lower the uncertainty of strontium lattice clocks an interrogation of the atoms in a colder environment is a viable solution \cite{mid11}. E.g. for a $1\times10^{-18}$ relative frequency uncertainty, at liquid nitrogen temperature only an uncertainty of the temperature of about 1~K is needed, while the shift coefficient introduces a completely negligible uncertainty. 

\begin{acknowledgments}
We thank H.H.~Ernst, A.~Felgner, P.~Franke, D.~Hagedorn, R.~Kr\"uger-Sehm, F.~Lechelt, A.~Linkogel, and M.~Schulz for valuable technical assistance and R.~Zirpel for equipment loan. This work was financially supported by the European Metrology Research Programme EMRP under IND 014 and the Centre for Quantum Engineering and Space-Time Research QUEST. The EMRP is jointly funded by the EMRP participating countries within EURAMET and the European Union.
\end{acknowledgments}

%


\section{SUPPLEMENTARY MATERIAL}
This supplement covers how we obtain the dynamic correction and an estimate of additional corrections that are needed to apply  our measurement on the isotope $^{88}$Sr also to $^{87}$Sr.

\section{Fitting Procedure}
The dynamic correction $\Delta\nu_{\rm dyn}$ of the blackbody radiation (BBR) shift depends on the transition probabilities connecting the clock states to other states. In strontium it is mostly determined by the strength of the transition with the longest wavelength $5s4d~^3$D$ \rightarrow 5s5p~^3$P$_0$ at 2.6~$\mu$m. 
Improved values of this transition amplitude as well as of other relevant transitions are obtained from a least square fit.

In the fit we use a dataset of experimental values like atomic life times, the magic wavelength or the clock laser light shift in addition to the differential polarizability measured in this work. No correlation between the data was assumed as it originates from various unrelated studies. 
Based on Einstein $A$ coefficients and transition frequencies we calculate these observables.
Here we use the electric dipole approximation. 
Contributions of $E2$ and $M1$ transitions contribute with $2.4\times10^{-5}$~Hz and $2.4\times10^{-8}$~Hz respectively to the room temperature blackbody radiation shift \cite{por06} and can be neglected on the level of uncertainty considered here.
In the fit procedure, the values of selected $A$ coefficients (Type F in Tab.~\ref{tab:fit}) were adjusted to achieve the best overall agreement with the dataset.

For the transitions that most contribute to the polarizability, experimental line strength data is available which was also included in the dataset of observables (Type D in Tab.~\ref{tab:fit}). 
This results in a soft constraint of these free parameters. 
For all lines (Tables \ref{tab:lines_1S0} and \ref{tab:lines_3P0}) we have used experimental uncertainties to the $A$ coefficients as given in the references. 
To the aforementioned 2.6~$\mu$m line we have assigned an increased uncertainty due to the inconsistent literature values \cite{por08, mil92} to avoid a too strong constraint in the dynamic correction, which is strongly correlated to this value.

The contributions of transitions to high lying $ns$ and $nd$ states and of the continuum were estimated using hydrogen wave functions as in \cite{opp69}. To these contributions an average transition energy corresponding to the energy difference to the ionization limit was assigned. 

Choosing other lines as free parameters in the fit (marked by an asterisk in Tables~\ref{tab:lines_1S0} and \ref{tab:lines_3P0}) did neither reduce the residuals significantly nor change the final result for the dynamic correction. 

The Thomas-Reiche-Kuhn (TRK) sums for the $^1$S and $^3$P systems were mainly used to avoid extreme excursions of single oscillator strengths that are not well constrained by the experimental data. 
Deviations from the TRK-rule are expected from relativistic corrections \cite{lev57}. In the strontium spectrum relativistic corrections are on a fractional level of $5\%$ of the term energies \cite{dzu03}, which was also assumed as uncertainty in the TRK sum.

From the magic wavelengths at 813~nm and 390~nm only the first one was included in the fit. 
The 390-nm wavelength depends on details in the spectral region not of interest here. 
The uncertainty of the 813-nm wavelength was increased to 1.5~GHz to allow for isotope shift of its value and corrections due to two photon transitions (hyperpolarizability) \cite{wes11}. 
\begin{figure}[t]
\includegraphics[width=\columnwidth]{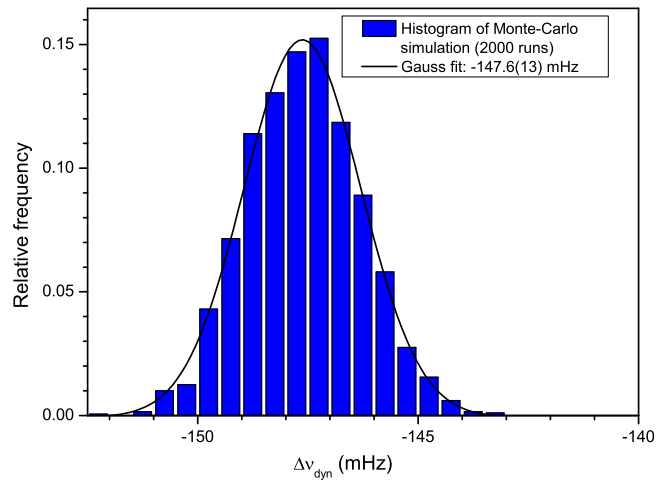}	
\caption{Probability distribution of the dynamic BBR shift $\Delta \nu_{\rm dyn}(300~{\rm K})$ together with a Gaussian fit to determine its value and the associated uncertainty.}
\label{fig:dyn}
\end{figure}
For the static polarizability of the ground state we use an experimental value \cite{sch74}. For comparison we include in our calculation an additional contribution of the atomic core as in \cite{por08}. This contribution cancels for all other observables, as it is the same for both states. 

The fitted parameters are shown in Tab.~\ref{tab:fit}. Overall a good agreement with the literature values is achieved, indicating a consistent description.

To estimate an uncertainty of the value of $\Delta \nu_{\rm dyn}$ that can be calculated from the fitted $A$ coefficients, a Monte-Carlo simulation is performed. 
Artificial datasets were generated by randomly choosing the experimental observables within their uncertainties. 
For each dataset a fit is performed and $\Delta \nu_{\rm dyn}$ is calculated. The probability distribution of $\Delta \nu_{\rm dyn}$ is shown in Fig.~\ref{fig:dyn} together with a Gaussian fit to determine the final value and its uncertainty.
Additionally we varied $A$ coefficients not included in the fit by up to 15\% to account for their uncertainties. We observed shifts of $\Delta \nu_{\rm dyn}$ of up to 1.0~mHz, which we added to the uncertainty from the Monte-Carlo simulation, resulting in
$\Delta \nu_{\rm dyn} = 147.6(23)~{\rm mHz}$.

\renewcommand{\arraystretch}{1.5}

\begin{table*}[t]
\begin{tabular}{| c | l | c | r | l | }
\hline
 Type & property & fit result & literature value & reference\\
\hline
 DF &$A (5s5p~^1$P$_1 \rightarrow 5s^2~^1$S$_0)$ at $\lambda = 461~$nm  			& $-0.017~\sigma$ 
	& 1.9001(14) $\times10^{8}$  & \cite{yas06} \\

 F &$A (4s5d~^1$P$_1 \rightarrow 5s^2~^1$S$_0)$ at $\lambda = 231~$nm  			& 6.6 $\times10^{6}~{\rm s}^{-1}$ 
	& $1.7 \times 10^{7}~{\rm s}^{-1}$ & \cite{par76}\\

\hline

 DF &$A (5s4d~^3$D$_1 \rightarrow 5s5p~^3$P$_0$) at 2.6~$\mu$m  	& $2.67 \times10^{5}~{\rm s}^{-1} (-0.3~\sigma)$ 
	& $2.9(8)\times10^{5}~{\rm s}^{-1}$ 
	& see text \\

 DF &$A (5s6s~^3$S$_1 \rightarrow 5s5p~^3$P$_0)$ at 679~nm  			& 7.55 $\times10^6~{\rm s}^{-1} (-2.7~\sigma)$ 
	&  8.9(5)$\times10^6~{\rm s}^{-1}$ 
	& \cite{joe84} \\

 F &$A (5s5d~^3$D$_1 \rightarrow 5s5p~^3$P$_0)$ at 483~nm  			&  $3.34\times10^7~{\rm s}^{-1}$  
	&  3.3(2) $\times10^7~{\rm s}^{-1}$ 
	& \cite{gar88, san10a} \\

 F &$A_{\rm eff} (^3$D$_1$ continuum$ \rightarrow^3$P$_0)$ at $\lambda \approx 316~$nm  &  $9.71\times10^7~{\rm s}^{-1}$
	&  5.6 $\times10^{7}~{\rm s}^{-1}$ 
	& \cite{opp71} \\

\hline
 D &$5s^2~^1$S$_0$ Thomas-Reiche-Kuhn sum 			&  1.97  	
	&  2.0(1)  	& \cite{con53} \\	

 D &$5s5p~^3$P$_0$ Thomas-Reiche-Kuhn sum 			&  1.99  	
	&  2.0(1)  	& \cite{con53} \\	

 D & $\Delta\alpha$($5s^2~^1$S$_0$ - $5s5p~^3$P$_0$) & $-9\times10^{-5}~\sigma$
	& $4.07873(11)\times10^{-39}~{\rm Cm}^2/{\rm V}$ & this work \\

 D &static polarizability $5s^2 ~^1$S$_0$  				& $0.6~\sigma$
	&  $3.1(3)\times10^{-39}~{\rm Cm}^2/{\rm V}$ 	& \cite{sch74} \\

 D &$d\alpha/d\nu$ at 813 nm magic wavelength &	$-2.6~\sigma$
	& $2.2(2)\times10^{-5}{\rm~Hz}/(E_{\rm r}\cdot {\rm MHz})$ 
	& \cite{let07a} \\

 D &magic wavelength near 813 nm								& $3\times10^{-4}~\sigma$
	& 368554.7(15)~GHz	& \cite{wes11}, see text\\

 D &dynamic polarizability at clock transition& 	$-5.4~\sigma$ 
	&-13(2)~Hz/(W\,cm$^{-2})$ & \cite{bai07} \\
	
 u & magic wavelength near 390 nm								& 	767344~GHz 		
	& 768917(18)~GHz	& \cite{tak09} \\
\hline 
\end{tabular}
\caption{Input data and fit results in the determination of the dynamic BBR shift. 
Type D denotes experimental data used in the least square fit, F denotes atomic parameters that are varied during the fit and u denotes experimental data that is not included in the fit.}
\label{tab:fit}
\end{table*}
\begin{table*}[t]
\begin{tabular}{| c | c | c | cc | }
\hline
upper level& wavelength & $A_{ki}$ (s$^{-1}$) & \multicolumn{2}{c|}{reference}\\
\hline
$5s5p~^3$P$_1$ & 	689 nm 	& $4.69(11)\times10^4$		& \cite{dro97} & \\
$5s5p~^1$P$_1$ &	461 nm 	& $1.9001(14)\times10^8$	& \cite{yas06} & F\\
$5s6p~^1$P$_1$ &  293 nm 	& $1.9\times10^6$ 		& \cite{par76} & \\	
$5s7p~^1$P$_1$ &  257 nm 	& $5.3\times10^6$	 		& \cite{par76} & \\	
$5s8p~^1$P$_1$ &	243 nm 	& $1.8\times10^7$	 		& \cite{par76} & \\	
$5s9p~^1$P$_1$ &	236 nm 	& $1.15\times10^7$		& \cite{par76} & \\	
$5s10p~^1$P$_1$& 	231 nm 	& $6.6\times10^6$	 		& \cite{par76} & F \\	
$4d5p~^1$P$_1$ &	243 nm 	& $1.7\times10^7$			& \cite{par76} & \\	
Rydberg and cont.$~^1$P$_1$ &	$\approx 218$~nm 	& $\approx~10^7$			& estimated from \cite{gar83} & *\\	
\hline
\end{tabular}
\caption{Lines connecting to the $5s^2~^1$S$_0$ level. Lines used in the fit are marked by F, asterisks denote lines that have been used as alternative free parameters. }
\label{tab:lines_1S0}
\end{table*}

\begin{table*}[t]
\begin{tabular}{| c | c | c | cc | }
\hline
upper level& wavelength & $A_{ki}$ (s$^{-1}$) &  \multicolumn{2}{c|}{reference}\\
\hline
$5s6s~^3$S$_1$ & 679 nm & $8.9(5)\times10^6$	& \cite{joe84} & F\\
$5s7s~^3$S$_1$ & 433 nm & $3.1(2)\times10^6$	& \cite{gar88, san10a} & \\
$5s8s~^3$S$_1$ & 378 nm & $9.1 \times10^5$		& \cite{wer92} & \\
$5s9s~^3$S$_1$ & 355 nm & $5.0 \times10^5$		& \cite{wer92} & \\
$5s10s~^3$S$_1$& 344 nm & $3.1 \times10^5$		& \cite{wer92} & \\

$5p^2~^3$P$_1$ &	 474 nm & $3.9(3)\times10^7$	&	\cite{gar88, san10a} & * \\
$4d^2~^3$P$_1$ &	 330 nm & $5.9(9)\times10^7$	&	\cite{kel76, san10a} & * \\

$5s4d~^3$D$_1$&	2603 nm & $2.9(8) \times10^5$	& see text & F \\
$5s5d~^3$D$_1$&	 483 nm & $3.3(2)\times10^7$	&	\cite{gar88, san10a} & F \\
$5s6d~^3$D$_1$&	 394 nm & $1.48  \times10^7$	& \cite{por08} & \\
$5s7d~^3$D$_1$&	 363 nm & $7.9   \times10^6$	& \cite{wer92} & \\
$5s8d~^3$D$_1$&	 348 nm & $4.7   \times10^6$	& \cite{wer92} & \\
$5s9d~^3$D$_1$&	 339 nm & $3.1   \times10^6$	& \cite{wer92} & \\

Rydberg and cont.$~^3$S$_1$&	 $\approx 316$~nm & $1.1\times10^7$	& \cite{opp69} & * \\
Rydberg and cont.$~^3$D$_1$& 	 $\approx 316$~nm & $5.6\times10^7$	& \cite{opp69} & F \\
\hline
\end{tabular}
\caption{Lines connecting to the $5s5p~^3$P$_0$ level. Lines used in the fit are marked by F, asterisks denote lines that have been used as alternative free parameters. }
\label{tab:lines_3P0}
\end{table*}

\section{isotope effects}
In order to apply our results obtained with $^{88}$Sr to optical clocks operating with the fermionic isotope $^{87}$Sr we repeat our calculation, taking into account the hyperfine structure due to the nuclear spin $I=9/2$ and the isotope shift.  

We rewrite the dynamic polarizability as 
\begin{equation}
\alpha_i(\omega) = \frac{4\pi}{h}
\sum_{k}  \frac{\omega_{ik} d_{ik}^2}{(\omega_{ik}^2-\omega^2)}
\label{eq:alphaS}
\end{equation}
with transition dipole elements $d_{ik}^2 = \frac{\epsilon_0 c^3 h}{2} \frac{2 J_k+1}{2 J_i+1} A_{ki}/\omega_{ik}^3$ for a transition with spontaneous emission rate $A_{ki}$ from state $k$ to $i$ with angular momentum $J_k$ and $J_i$.

In a nonrelativistic approximation, the finite nuclear mass $M$ is taken into account by replacing the electron mass $m_e$ in the Schr\"odinger equation with the reduced mass 
\begin{equation} 
\mu = \frac{M m_{\rm e}}{M + m_{\rm e}}.
\end{equation}
Together with the electron mass also the atomic units scale, e.g. for a finite nuclear mass the atomic unit of lengths $a_0$  is replaced by
\begin{equation} 
a = \frac{\mu}{m_{\rm e}} a_0.
\end{equation}
Thus, in this approximation the dipole matrix elements $d_{ki}$ scale proportionally to the reduced mass as well \cite{wie09a}. Comparing $^{88}$Sr to $^{87}$Sr, this amount amounts to 
\begin{equation} 
\frac{^{87}d_{ik}}{^{88}d_{ik}} = \frac{^{88}a}{^{87}a} = 1 + 7\times10^{-8}.
\end{equation} 
  
If this scaling would also apply to the transition energies (i.e. normal mass shift $\omega_{ik}\propto \mu$), the polarizability would scale to the third power of the ratio of the reduced masses and the dynamic correction to the fifth power \cite{mid11}. 
However, in addition there are contributions to the isotope shift from electron correlations (specific mass shift) and nuclear size (volume shift). 
Thus, we use available experimental data for the transition shifts to calculate the difference between the two isotopes, similar as in high accuracy calculations of the helium \cite{pac00,cen01,gai08} or lithium \cite{tan09} polarizability.
For lines where no data was available we have assumed an isotope shift of 150 MHz, which is larger than any observed shift.

Where available, we have included hyperfine splittings. As the hyperfine interaction leaves the center of gravity unchanged in first order it has no influence on the dynamic polarizability (Eq. \ref{eq:alphaS}). 
Even with large hyperfine splitting on the prominent line $^3$P$_0$ -- $^3$D$_1$ \cite{bus93}, it leads to a correction of less than $10^{-6}$. Thus no hyperfine splitting was used for lines where no data was available. 

From our calculation we obtain a change of the BBR shift of
\begin{equation}
^{88}\Delta\nu_{\rm stat} - {^{87}\Delta\nu_{\rm stat}}= 2.98~\mu{\rm Hz}, 
\end{equation} 
or $1.3 \times 10^{-6}$ of the total BBR shift. 
The corresponding change of the dynamic correction amounts to 
\begin{equation}
^{88}\Delta\nu_{\rm dyn} - {^{87}\Delta\nu_{\rm dyn}} =0.5~\mu{\rm Hz},
\end{equation} 
or $3.4 \times 10^{-6}$ of the dynamic correction.
Thus on the level of accuracy that we have achieved, to $^{87}$Sr and $^{88}$Sr the same BBR correction can be applied.  

\end{document}